\documentclass[a4paper]{article}
\usepackage{url}
\usepackage{amsfonts,amsmath}
\usepackage{graphicx}
\begin{document}

\title{Tuning Technique for Multiple Precision Dense Matrix Multiplication using Prediction of Computational Time}
\author{Tomonori Kouya\thanks{\tt kouya.tomonori\@sist.ac.jp}\\Shizuoka Institute of Science and Technology}
\date{\today}
\maketitle
\begin{abstract}
Although reliable long precision floating-point arithmetic libraries such as QD and MPFR/GMP are necessary to solve ill-conditioned problems in numerical simulation, long precision BLAS-level computation such as matrix multiplication has not been fully optimized because tuning costs are very high compared to IEEE float and double precision arithmetic. In this study, we develop a technique to shorten this tuning time by using prediction of computational times in several block sizes for the blocking algorithm, and then selecting the fastest matrix multiplication method for tuning multiple precision dense real matrix multiplication in various precisions, matrix sizes, and degrees of parallelization.
\end{abstract}

\section{Introduction}

As scientific computations increase in size, so-called ``ill-conditioned problems" have emerged in various fields. These are problems whose numerical results cannot be validated by using IEEE754 double precision floating-point arithmetic. Reliable software-based fixed and arbitrary precision floating-point arithmetic libraries such as QD(Double-double and Quadruple-double precision) library developed by Bailey et.al\cite{qd} and MPFR\cite{mpfr}/GMP\cite{gmp}, are very effective tools for these ill-conditioned problems. But we cannot use optimized BLAS-level multiple precision computations. 

We therefore have implemented three types of dense matrix multiplication, such as simple triple loop (simple for short), blocking (block for short), and Strassen's algorithms (Strassen for short) in QD and MPFR/GMP environments\cite{kouya_strassen2014}, and also parallelized them with OpenMP. By benchmarking them, we find that simple or blocking algorithms are more effective for small-sized and relatively low precision matrix multiplication, and that Strassen algorithm is the most effective for large-sized and high precision matrix multiplication\cite{kouya_strassen2016}. As a result, we have developed our original dense matrix multiplication library to effectively compute multiple precision dense matrix products by using an automatic tuning method in order to select the most effective algorithm. It also involves a technique to shorten tuning time by using prediction of computational times in several block sizes for the block algorithm.

In this paper, we explain our implemented prediction of computational times for a blocking algorithm, and then show its effectiveness for tuning multiple precision dense matrix multiplication in various precisions, matrix sizes, and degrees of parallelization.

\section{Implemented dense matrix multiplication algorithms}
As mentioned previously, we have already implemented three types of algorithms for dense real matrix multiplication with QD and MPFR/GMP. We consider a dense real matrix multiplication to obtain $C := AB\in\mathbb{R}^{m\times n}$, where $A =[a_{ij}]$ $\in \mathbb{R}^{m\times l}$, $B = [b_{ij}]$$\in \mathbb{R}^{l\times n}$ and $c_{ij}$, the elements of $C$ are obtained as 
\begin{equation}
	c_{ij} := \sum^l_{k = 1} a_{ik} b_{kj}. \label{eqn:matrix_mul_simple}
\end{equation}
A calculation following the above definition (\ref{eqn:matrix_mul_simple}) is called a ``simple" algorithm.

To draw the power of cache memories in the majority of current CPUs, ``block" algorithms are used in dense matrix multiplications, which normally divide $A$ and $B$ into small blocked matrices with the number of their rows and columns less than the block size $n_{\rm min}$. For example, if $A$ are divided to $M\times L$ small matrices $A_{ik}$ and $B$ to $L\times N$ small matrices $B_{kj}$, then $C$ are calculated with $M\times N$ small matrices $C_{ij}$, which are
\begin{equation}
	C_{ij} := \sum^L_{k = 1} A_{ik} B_{kj}. \label{eqn:matrix_mul_blocking}
\end{equation}

The ``Strassen" algorithm is one of divide-and-conquer methods. For even dimensional matrices $A$ and $B$, four divided matrices are prepared such that

\[
 A = \left[\begin{array}{cc}
	A_{11} & A_{12} \\
	A_{21} & A_{22}
\end{array}\right],\ B = \left[\begin{array}{cc}
	B_{11} & B_{12} \\
	B_{21} & B_{22}
\end{array}\right] 
\]
where $A_{ij} \in \mathbb{R}^{m/2\, \times\, l/2}$ and $B_{ij} \in \mathbb{R}^{l/2\, \times\, n/2}$．By using these block matrices, the Strassen algorithm performs the following recursive calculations to obtain $P_i$ $(i = 1, 2, ..., 7)$, and finally obtain $C$ with these $P_i$s.

\begin{equation}
\begin{split}
	P_1 &:= (A_{11} + A_{22}) (B_{11} + B_{22}),\ P_2 := (A_{21} + A_{22}) B_{11}, \\
	P_3 &:= A_{11} (B_{12} - B_{22}),\ P_4 := A_{22} (B_{21} - B_{11}), \\
	P_5 &:= (A_{11} + A_{12}) B_{22},\ P_6 := (A_{21} - A_{11}) (B_{11} + B_{12}), \\
	P_7 &:= (A_{12} - A_{22}) (B_{21} + B_{22}) \\
	C &:= \left[\begin{array}{cc}
		P_1 + P_4 - P_5 + P_7 & P_3 + P_5  \\
		P_2 + P_4 & P_1 + P_3 - P_2 + P_6
\end{array}\right]
\end{split} \label{eqn:matrix_mul_strassen}
\end{equation}

We have parallelized the above three algorithms (\ref{eqn:matrix_mul_simple}), (\ref{eqn:matrix_mul_blocking}), and (\ref{eqn:matrix_mul_strassen}) with OpenMP and confirmed their effectiveness\cite{kouya_strassen2016}.

\section{Prediction of computational times for block algorithm}

We must select the optimal block size for the block algorithm to draw the power of the cache memory in CPUs. In IEEE float and double precision, optimized BLAS libraries such as ATLAS\cite{atlas}, have a mechanism to select the most effective block sizes using trial-and-error methods. Recent hardware evolution requires such optimization for software-based multiple precision floating-point arithmetic such as performed by QD and MPFR/GMP. These floating-point numbers are implemented with complex data types. Therefore, we cannot uniformly select the optimal block sizes for various precisions and must also expend significant time to compute these multiple precision dense matrix multiplications.

\begin{figure}[ht]
\centering
\includegraphics[width=.7\textwidth]{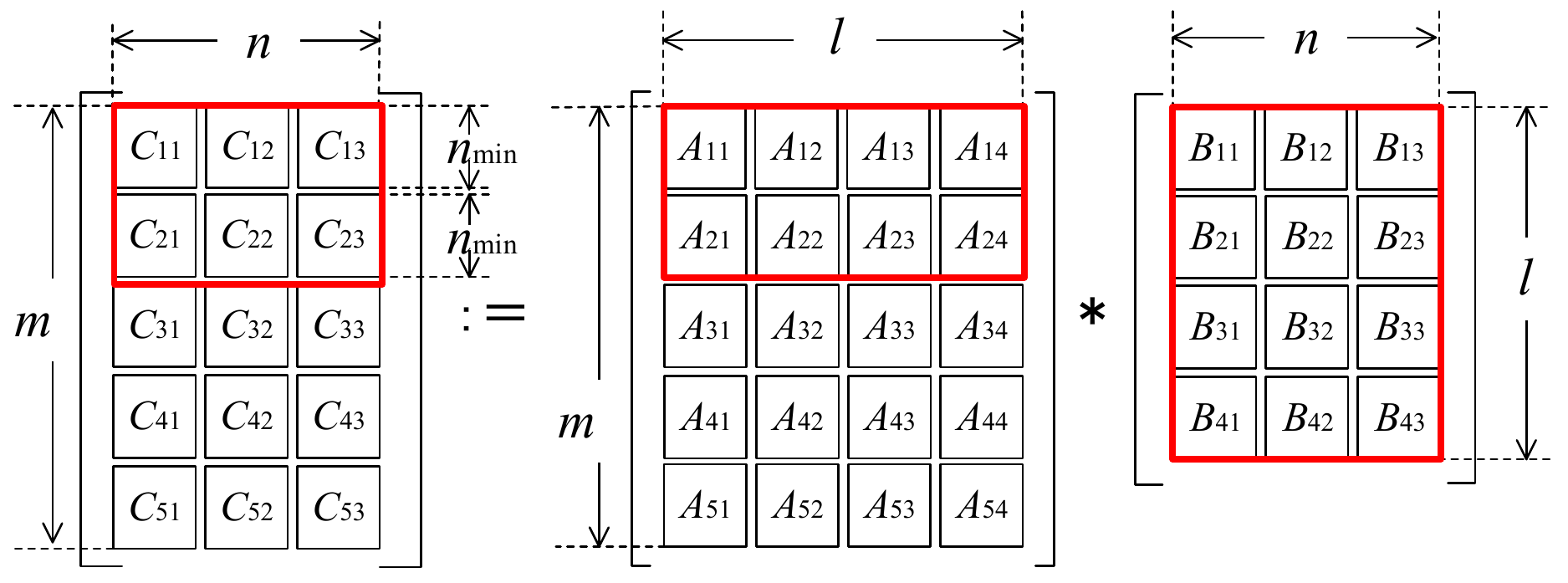}
\caption{The sections surrounded by red lines are used for prediction of computational times of full size block algorithms}\label{fig:block_algorithm}
\end{figure}

To shorten the optimization tuning time, we develop an original prediction method of computational times for a block algorithm. Our prediction technique is defined by the following process:
\begin{enumerate}
	\item Obtain the computational time using $A$ with shorter rows and original $B$ as shown in \figurename\ref{fig:block_algorithm}.
	\item Predict computational time of the full-size matrix multiplication by using the above computational time.
\end{enumerate}
By applying the above predictions for various block sizes, we can select the optimal block size $n_{\rm min}$.

Our prediction technique can obtain precise computational time because the hit ratio of caches is equivalent to full-size matrix multiplication using the original $B$. 

Our tuning program involving this prediction technique is executed as follows:
\begin{enumerate}
	\item Predict computational times of full sizes with several candidates of optimal block size and select the optimal $n_{\rm min}$.
	\item Determine the computational times of simple algorithm, the block algorithm with the optimal $n_{\rm min}$, and the Strassen algorithm.
	\item Select the fastest algorithm among three types of algorithms.
\end{enumerate}

The above tuning process would require 1.6 to 2.3 times higher tuning costs without our prediction technique in (1). 

\section{Accuracy of prediction of computational time and whole tuning results}

We use the dense real matrix multiplication $C := AB$ as follows:
\[ A = \left[\sqrt{5} \left(i + j - 1\right)\right]^{n}_{i,j=1},\ B = \left[\sqrt{3}\left(n - i\right)\right]^{n}_{i,j=1}. \]
The reason why we select these squared $A$ and $B$ matrices is that elementary arithmetic provided by MPFR/GMP uses an economical way if the elements of these matrices are not fulfilled in a binary mantissa. The same matrices are applied in DD and QD computation.

In this paper, the following three types of computational environments are tested. Three abbreviations for these environments are ``Xeon"，``Corei7", and ``Ryzen" associated with the names of each CPU. All are x86\_64 Linux boxes, but exhibit higher performance in ascending order.

\begin{description}
	\item[Xeon] Intel Xeon E5-2620 v2 (2.10GHz, 6 cores)×2, 32GB RAM, CentOS 6.5 x86\_64, Intel C/C++ 13.1.3, MPFR 3.1.2 / GMP 6.0.0a $+$ BNCpack 0.8\cite{bnc}
	\item[Corei7] Intel Core i7-6700K (4.00GHz, 4 cores)，16GB RAM，Ubuntu 16.04 x86\_64, GCC 5.4.0, QD, MPFR 3.1.6 / GMP 6.1.2 + BNCpack 0.8
	\item[Ryzen] AMD Ryzen 1700 (1.55Hz, 8 cores), 16GB RAM, Ubuntu 16.04 x86\_64, GCC 5.4.0, MPFR 3.1.5 / GMP 6.1.2 $+$ BNCpack 0.8
\end{description}

\subsection{Fixed precision matrix multiplication with QD}

In case of $n = 1024$, the prediction results with QD are shown in \tablename\ \ref{table:qd_est_xeon_ryzen}, where (a) means optimal block size, (b) is the real computational time (unit: second) of the corresponding block size, (c) is the computational time (unit: second) needed for the prediction, (d) is the predicted computational time (unit: second) of the full-size matrix multiplication, ``rel.diff" means relative difference of predicted computational time, and  (e) is the computational time  (unit: second) of the Strassen algorithm for reference. Underlined values indicate the minimal computational time with the corresponding precision and matrix size.

\begin{table}[ht]
\centering
\caption{Relative Error of Estimated Computational Times for DD and QD:\ Xeon(Upper)，Corei7(Middle), and Ryzen(Lower)}\label{table:qd_est_xeon_ryzen}
\begin{tabular}{|c|c|c|c|c|c|c|c|} \hline
prec.	& \#thr. & 	(a) & (b)& (c) & (d) & rel.diff. & (e) \\ \hline\hline
DD &	1&		64  &	45.97&	5.77 &	46.1&	0.28\%&	\underline{25.54} \\ \hline
DD &	2&		64&	23.26&	2.92&	23.3&	0.17\%&	\underline{14.75} \\ \hline
DD &	4&		64&	12.29&	1.54&	12.3&	0.08\%&	\underline{7.779} \\ \hline
DD &	8&		64&	6.518&	0.809&	6.47&	0.74\%&	\underline{4.053} \\ \hline\hline
QD &	1&		512&	491.5&	490&	490&	0.31\%&	\underline{270.3} \\ \hline
QD &	2&		512&	248.5&	249&	249&	0.20\%&	\underline{155.3} \\ \hline
QD &	4&		128&	130.2&	32.5&	130&	0.15\%& \underline{81.10}  \\ \hline
QD &	8&		128&	67.27&	16.8&	67.3&	0.04\%&	\underline{42.06} \\ \hline
\end{tabular}
								
\begin{tabular}{|c|c|c|c|c|c|c|c|} \hline
prec.	& \#thr. & 	(a) & (b)& (c) & (d) & rel.diff. & (e) \\ \hline\hline
DD&	1&		512&	26.13&	26.1&	26.1&	0.11\%&	\underline{15.29} \\ \hline
DD&	2&		256&	13.12&	6.56&	13.1&	0.15\%&	\underline{8.833} \\ \hline
DD&	4&		256&	6.959&	3.48&	6.96&	0.01\%&	\underline{4.762} \\ \hline\hline
QD&	1&		256&	213.1&	107&	213&	0.05\%&	\underline{117.1} \\ \hline
QD&	2&		256&	106.7&	53.4&	107&	0.28\%&	\underline{66.97} \\ \hline
QD&	4&		128&	53.76&	13.4&	53.8&	0.07\%&	\underline{33.64} \\ \hline
\end{tabular}
								
\begin{tabular}{|c|c|c|c|c|c|c|c|} \hline
prec.	& \#thr. & 	(a) & (b)& (c) & (d) & rel.diff. & (e) \\ \hline\hline
DD&	1&		128&	12.57&	3.15 &	12.6&	0.24\% &\underline{8.112} \\ \hline
DD&	2&		256&	6.716&	3.36&	6.71&	0.09\%&	\underline{5.117} \\ \hline
DD&	4&		256&	4.064&	2.01&	4.01&	1.33\%&	\underline{3.024} \\ \hline
DD&	8&		128&	2.128&	0.554&	2.22&	4.32\%&	\underline{1.557} \\ \hline\hline
QD&	1&		256&	86.64&	43.4&	86.7&	0.07\%&	\underline{51.14} \\ \hline
QD&	2&		128&	45.08&	11.2&	44.9&	0.40\%&	\underline{29.50} \\ \hline
QD&	4&		256&	26.21&	13.2&	26.5&	1.11\%&	\underline{17.37} \\ \hline
QD&	8&		128&	14.66&	3.44&	13.8&	5.87\%&	\underline{8.940} \\ \hline
\end{tabular}

\end{table}

All relative differences between predicted and real computational times are under 6\%．We have confirmed that these relative differences for other dimensions are under 10\%, so our prediction technique is precise.

Next, we show the tuning results of simple algorithm, the block algorithm with the optimal block size, and the Strassen algorithm in \tablename\ \ref{table:dd_qd_tuning_time}. From our previous study\cite{kouya_strassen2014}\cite{kouya_strassen2016}, it is known that the Strassen algorithm is the most effective over a certain dimensional threshold, so total tuning time and the threshold are shown in \tablename\ \ref{table:dd_qd_tuning_time}.

\begin{table}[htbp]
\centering
\caption{Total Times for Tuning in DD and QD computations}\label{table:dd_qd_tuning_time}
\begin{tabular}{|c|c|c|c|} \hline
		& Xeon		& Corei7	& Ryzen \\ \hline
$n$		& \multicolumn{3}{|c|}{128($\pm 1$), 256($\pm 1$), 512($\pm 1$), 1024($\pm 1$)} \\ \hline
$n_{\rm min}$	& \multicolumn{3}{|c|}{8, 16, 32, 64, 128, 256, 512} \\	\hline
\# of Threads		& 1, 2, 4, 8 	& 1, 2, 4	& 1, 2, 4, 8 \\ \hline
Total Time(DD)	& 5.7 h 	& 4.1 h		& 3.4 h \\ 
Threshold (DD)	& $n \ge 64$	&  $n \ge 128$	& $n \ge 257$ \\ \hline
Total Time(QD)	& 34.0 h		& 14.3 h	& 11.3 h \\ 
Threshold (QD)	& $n \ge 63$	&  $n \ge 128$	& $n \ge 256$ \\ \hline
\end{tabular}
\end{table}

\tablename\ \ref{table:dd_qd_tuning_time} indicates that the more powerful computational environment we use, shorter tuning time and larger threshold are obtained.

\subsection{Arbitrary precision matrix multiplication with MPFR/GMP}
%

\tablename\ \ref{table:mpfr_relerr_est} shows the results obtained by using dense real matrix multiplication in case of $n = 1024$ with MPFR/GMP. Arbitrary precision is available in MPFR/GMP, so precision is shown only in cases of 128- and 1024-bit mantissas.

\begin{table}[ht]
\centering
\caption{Relative Error of Estimated Computational Times for MPFR/GMP:\ Xeon(Upper)，Corei7(Middle), and Ryzen(Lower)}\label{table:mpfr_relerr_est}
\begin{tabular}{|c|c|c|c|c|c|c|c|} \hline
prec.	& \#thr. & 	(a) & (b)& (c) & (d) & rel.diff. & (e) \\ \hline\hline
128 &	1&		32 &	131.4&	8.24&	132 &	0.46\%&	\underline{52.23}\\ \hline
128&	2&		32&	\underline{67.72} &	4.43&	70.8&	4.55\%&	88.15\\ \hline
128&	4&		32&	\underline{36.70} &	2.38&	38.1&	3.81\%&	70.38\\ \hline
128&	8&		32&	\underline{19.75} &	1.31&	21&	6.33\%&	51.64\\ \hline\hline
1024&	1&		32&	579.7&	36.2&	579&	0.12\%&	\underline{122.8}\\ \hline
1024&	2&		16&	349.7&	11.2&	359&	2.66\%&	\underline{171.2}\\ \hline
1024&	4&		128&	198.9&	51.6&	207&	4.07\%&	\underline{138.7}\\ \hline
1024&	8&		16&	170.8&	5.3&	169&	1.05\%&	\underline{97.70} \\ \hline
\end{tabular}
								
\begin{tabular}{|c|c|c|c|c|c|c|c|} \hline
prec.	& \#thr. & 	(a) & (b)& (c) & (d) & rel.diff. & (e) \\ \hline\hline
128  &	1&		64 &	83.45&	10.2&	82  &	1.74\%	&\underline{24.88} \\ \hline
128&	2&		32&	41.77&	2.63&	42&	0.55\%	&\underline{31.21} \\ \hline
128&	4&		32&	\underline{21.51} &	1.35&	21.6&	0.42\%	&27.23 \\ \hline\hline
1024&	1&		64&	306.6&	38.2&	306&	0.20\%	&\underline{82.40}  \\ \hline
1024&	2&		16&	182.7&	5.75&	184&	0.71\%	&\underline{68.43} \\ \hline
1024&	4&		128&	97.31&	24.5&	98&	0.71\%	&\underline{43.63} \\ \hline
\end{tabular}
								
\begin{tabular}{|c|c|c|c|c|c|c|c|} \hline
prec.	& \#thr. & 	(a) & (b)& (c) & (d) & rel.diff. & (e) \\ \hline\hline
128 &	1&		128&	84.18&	21.1&	84.4&	0.26\%&	\underline{33.96}\\ \hline
128&	2&		128&	\underline{42.45}&	10.6&	42.5&	0.12\%&	64   \\ \hline
128&	4&		128&	\underline{25.13}&	6.43&	25.7&	2.27\%&	48.99\\ \hline
128&	8&		128&	\underline{13.02}&	3.33&	13.3&	2.15\%&	31.1 \\ \hline\hline
1024&	1&		32&	330.8&	20.7&	331&	0.06\%&	\underline{112.4} \\ \hline
1024&	2&		32&	217&	13.7&	219&	0.92\%&	\underline{131.2} \\ \hline
1024&	4&		32&	132.5&	8.23&	132&	0.38\%&	\underline{89.74} \\ \hline
1024&	8&		64&	70.65&	9.2&	73.6&	4.18\%&	\underline{54.41} \\ \hline
\end{tabular}

\end{table}

In this case, we confirmed that relative differences between predicted and real computational times are also within 7\%, so our prediction method is precise. We cannot find other cases in which relative differences are greater than 10\%.

\tablename\ \ref{table:mpfr_tuning_time} shows total tuning time of matrix multiplication with MPFR/GMP. MPFR/GMP provides arbitrary precision floating-point arithmetic, so longer tuning times are necessary because the number of trial patterns using 128 to 2048 bits is much larger than QD.

\begin{table}[htbp]
\centering
\caption{Total Times for Tuning in MPFR/GMP computation}\label{table:mpfr_tuning_time}
\begin{tabular}{|c|c|c|c|} \hline
		& Xeon		& Corei7	& Ryzen \\ \hline
$n$		& \multicolumn{3}{|c|}{128($\pm 1$), 256($\pm 1$), 512($\pm 1$), 1024($\pm 1$)} \\ \hline
$n_{\rm min}$	& \multicolumn{3}{|c|}{8, 16, 32, 64, 128} \\	\hline
Precision	& \multicolumn{3}{|c|}{128, 256, 512, 1024, 2048} \\	\hline
\# of Threads	& 1, 2, 4, 8 	& 1, 2, 4	& 1, 2, 4, 8 \\ \hline
Total Time(MPFR)& 97.3 h	& 67.8 h		& 42.1 h \\ \hline
\end{tabular}
\end{table}

\begin{figure}[ht]
\centering
\includegraphics[width=.49\textwidth]{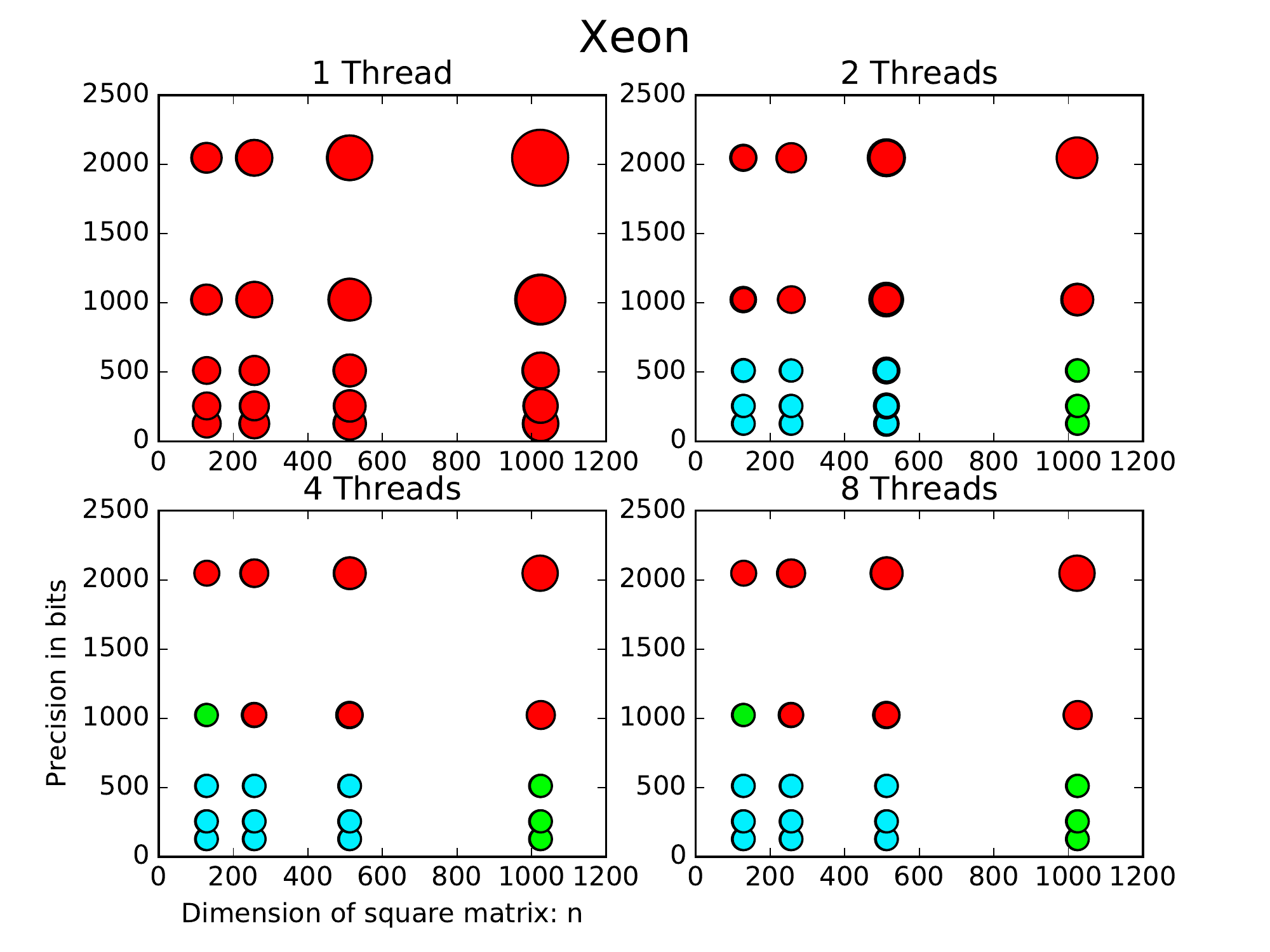}
\includegraphics[width=.49\textwidth]{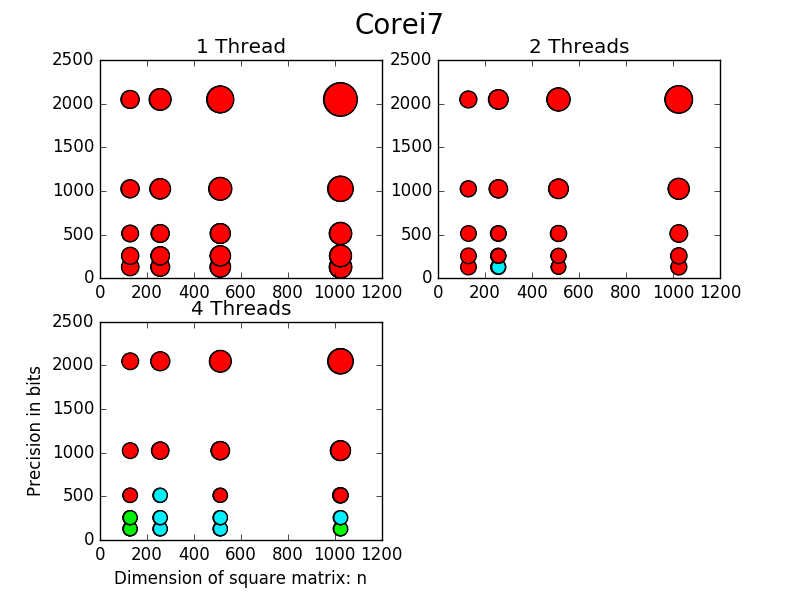}
\includegraphics[width=.49\textwidth]{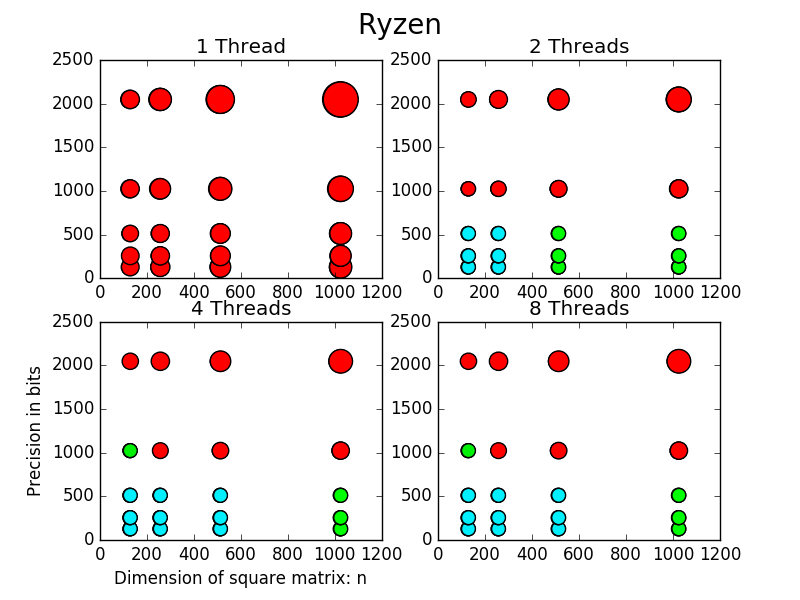}
\caption{Fastest Algorithms for all dimensions(horizontal line) and precisions(vertical line): Xeon(Upper)，Corei7(Middle)，and Ryzen(Lower)}\label{fig:mpfr_tuning_result}
\end{figure}

\figurename\ref{fig:mpfr_tuning_result} shows all tuning results obtained by using MPFR/GMP. Red circles in this figure indicate that the Strassen algorithm is the most effective，green circles indicate that block algorithm with 128 dimensional block size is the most effective, and other circles indicate that block algorithms with other block sizes are the most effective. Larger radiuses of circles indicate that they are more effective than other algorithms.

As mentioned in our previous study\cite{kouya_strassen2016}, our implemented parallelized Strassen algorithm is less effective for MPFR/GMP than for QD. The reason is that floating-point data types of MPFR are implemented with more complex structures, including pointers to the mantissa, compared to DD and QD.

Therefore, the more threads we use, the more cases in which the block algorithm is less effective. In fact, \figurename\ref{fig:mpfr_tuning_result} shows that the number of cases where block algorithms are more effective than the Strassen algorithm increases in any environment with less precision and more threads.

\begin{figure}[ht]
\centering
\includegraphics[width=.6\textwidth]{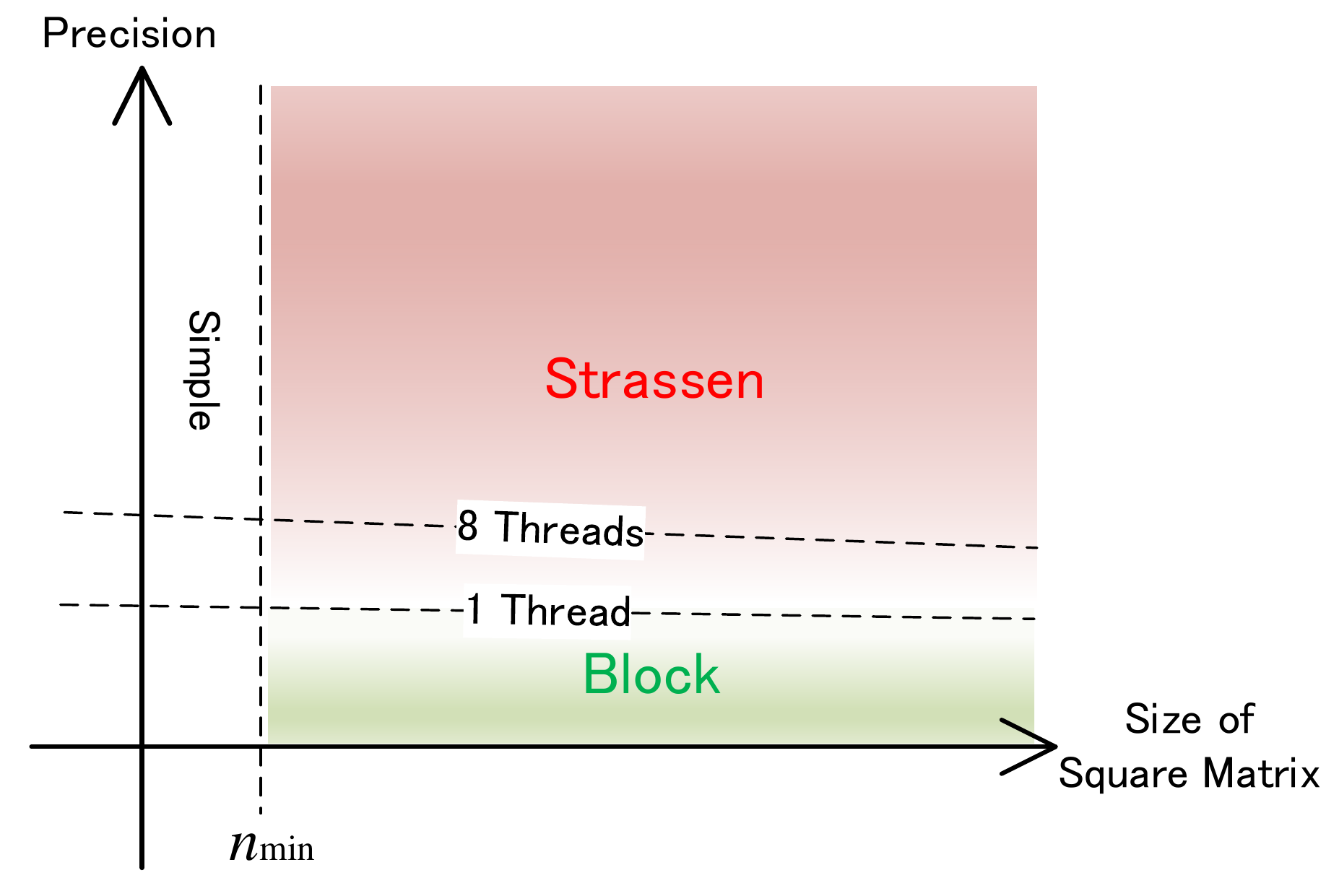}
\caption{The trend of the most effective algorithms with MPFR/GMP}\label{fig:mpfr_tuning_result_overall}
\end{figure}

\figurename \ref{fig:mpfr_tuning_result_overall} shows the general trend of tuning results in MPFR/GMP. Currently, the parallelized Strassen algorithm is not very effective, so the horizontal threshold where block algorithms are more effective than the Strassen algorithm slopes upward as the number of threads increases.

If users wish to optimize matrix multiplication around a wide threshold, they must run our tuning program with fixed dimensions of for targeted precision and matrix.

\section{Conclusion and future work}
By introducing a prediction technique for the block algorithm, we can automatically optimize dense real multiple precision matrix multiplication with QD and MPFR/GMP in shorter time.

Our future work will be to select appropriate precision and optimize matrix multiplication according to users' accuracy requirements.


\end{document}